\documentclass[draft]{aipproc}

\layoutstyle{6x9}

\begin{document}

\title{Analysis of Burst Observations by GLAST's LAT Detector}

\author{David L. Band}{
  address={Code 661, NASA/Goddard Space Flight Center, Greenbelt, MD 20771}}

\author{Seth W. Digel}{
  address={Stanford University, Stanford, CA}}

\author{the GLAST LAT Collaboration}{
  address={NASA-DOE-International Collaboration}}

\author{the GLAST Science Support Center (GSSC)}{
  address={Code 661, NASA/Goddard Space Flight Center, Greenbelt, MD 20771}}

\begin{abstract}
Analyzing data from GLAST's Large Area Telescope (LAT) will
require sophisticated techniques.  The PSF and effective
area are functions of both photon energy and the position
in the field-of-view. During most of the mission the
observatory will survey the sky continuously and thus the
LAT will detect each count from a source at a different
detector orientation; each count requires its own response
function!  The likelihood as a function of celestial
position and photon energy will be the foundation of the
standard analysis techniques.  However the 20 MeV--300 GeV
emission at the time of the $\sim$100 keV burst emission
(timescale of $\sim$10 s) can be isolated and analyzed
because essentially no non-burst counts are expected within
a PSF radius of the burst location during the burst.  Both
binned and unbinned (in energy) spectral fitting will be
possible. Longer timescale afterglow emission will require
the likelihood analysis that will be used for persistent
sources.
\end{abstract}

\maketitle


\section{Introduction}
The detection of the high energy emission from gamma-ray
bursts is anticipated to be one of the spectacular
observations by the Gamma-ray Large Area Space Telescope
(GLAST), NASA's next general gamma-ray astrophysics
mission.  Scheduled to be launched into low Earth orbit in
February, 2007, for 5--10 years of operation, GLAST will
consist of two instruments: the Large Area Telescope (LAT)
and the GLAST Burst Monitor (GBM).

A product of a NASA/DOE/international collaboration, the
LAT builds on the success of {\it CGRO}'s EGRET.  The LAT
will be a pair conversion telescope: gamma rays will
pair-produce in tungsten foils; silicon strip detectors
will track the resulting pairs; the resulting particles
will deposit energy in a CsI calorimeter; and an
anticoincidence detector will veto charged particles.  The
anticoincidence detector will be segmented to limit the
self-vetoing that plagued EGRET. The LAT will be 1.8
m$\times$1.8 m$\times$1m, and weigh $\sim$3000 kg.

The astrophysical photons will be only a small fraction of
the total number of events detected by the LAT, most of
which will result from charged particles.  On board
filtering of the events will reduce the $\sim$4 kHz trigger
rate to the $\sim$30 Hz event rate that can be downlinked
to the ground; ground processing will result in a $\sim$2
Hz photon rate.

The salient detector characteristics are: energy range of
$<$20 MeV to $>$300~GeV; 1--10~GeV effective area of
$>$8000~cm$^2$ with half maximum at 55$^\circ$; angular
resolution of $<3.5^\circ$ at 100~MeV, $<0.15^\circ$ at
10~GeV; field-of-view of $>$2~sr; deadtime $\sim20 \mu$s
per event (current, $<100 \mu$s required); and time
resolution of $\sim 2 \mu$s.

A descendant of CGRO's BATSE, the GBM will detect gamma-ray
bursts and extend GLAST's burst spectral sensitivity to the
$<$10 keV to $>$25 MeV band. Consisting of 12 NaI(Tl)
(10--1000 keV) and 2 BGO (0.15--25 MeV) detectors, the GBM
will monitor $>$8~sr of the sky, including the LAT's
field-of-view (FOV).  Bursts will be localized to
$<15^\circ$ (1$\sigma$) by comparing the rates in different
detectors.

Typically GLAST will survey the sky continuously.  After a
$\sim60$ day checkout phase, GLAST will undertake a one
year sky survey while the LAT team calibrates the
instrument. In survey mode GLAST will rock $\sim35^\circ$
above and below the orbital plane about the zenith
direction once per orbit. While pointed observations
proposed by guest investigators will be feasible during
subsequent years, continued survey mode operation will
usually be most efficient, and is expected to predominate.
Therefore most persistent sources will be observed at a
variety of detector orientations; each count will be
characterized by a different response function.

Both the GBM and the LAT will have burst triggers.  The GBM
will notify the LAT when it triggers.  When either
instrument triggers, a notice with a preliminary
localization will be sent immediately to the ground through
TDRSS and will then be disseminated by GCN. Additional data
will be downlinked through TDRSS for an improved
localization at the Mission Operations Center. Both
Instrument Operations Centers will calculate ``final''
positions from the full downlinked data.  GCN will
disseminate all positions.

The LAT will determine whether the burst was intense enough
for a followup 5~hour pointed observation at the burst
location (interrupted by earth occultations).  The
threshold will be higher for bursts the GBM detected
outside the LAT's FOV.

\section{Standard Source Analysis}

The LAT PSF is large ($\sim3.5^\circ$) at low energy
($\sim$100 MeV), small ($<0.15^\circ$) at high energy
($\sim$10 GeV). With the LAT's large effective area, many
sources will be detected; their PSFs will merge at low
energy. Therefore the analysis must be
3~dimensional---2~spatial and 1~spectral---and time is an
additional dimension for variable sources. Diffuse emission
underlies the point sources. For a typical analysis the
source model must include: all point sources within a few
PSF lengths of the region of interest; extended sources
(e.g., supernova remnants); spatially variable diffuse
Galactic emission (which must be modeled); and isotropic
extragalactic emission. Sources are defined by positions,
spectra, and perhaps time histories. Initial values may be
extracted from the point source catalog the LAT team will
compile. Consequently the source model will have many
parameters. In an analysis some will be fitted, some will
be fixed.

The instrument response (PSF, effective area, energy
resolution) will at least be a function of energy and angle
to the LAT normal; other parameters may be relevant such as
the azimuthal angle around the LAT normal or the conversion
layer (the front or back of the LAT).  Since the LAT will
usually survey the sky, a source will be observed at
different instrument orientations.  Each count will be
characterized by many observables, and therefore a very
large data space results. Even with $10^5$ counts, this
data space will be sparsely populated. Note that what high
energy astrophysicists call a ``count'' is a ``photon'' to
some particle physicists.

As was the case for EGRET\cite{1} and earlier gamma-ray
missions\cite{2}, likelihoods will be the foundation of our
analyses (e.g., detecting sources, determining source
intensities, fitting spectral parameters, setting upper
limits).  The likelihood is the probability of the data
(the counts that were detected) given the model (the photon
sources).  The data consist of both the counts that were
detected, and the regions of data space where counts were
not observed. The calculation of the likelihood will be
difficult because many counts will sparsely populate an
enormous data space.

The likelihood will be calculated many times as the source
model is changed (for example in fitting source
parameters), and factors that are not model-dependent
should be calculated once for a given analysis.  Many of
these quantities will have units of ``exposure''
(area$\times$time).
\section{Burst Spectral Analysis}
The duration of the $\sim$100 keV burst emission is
(relatively) short---at most 10's of seconds.  Therefore,
the LAT's pointing will not change significantly during the
burst, and all the counts can be treated as having one
response function. Within a PSF radius of the burst
position less than one non-burst count per minute is
expected:  [$\sim$2~Hz~cts over the FOV]\, /\, [2~sr FOV]
$\times$ [$\pi(3.5^\circ \pi/180)^2$~sr within PSF radius]
= $\sim$0.01~Hz~cts within a PSF radius or 0.7 cts/minute
within a PSF radius. Therefore, we can treat all counts
within 1--2 PSF radii as burst photons.

Since a)~all the counts within a PSF radius of the burst
originated in the burst, and b)~all the counts have the
same response function, multi-source spatial analysis is
unnecessary for spectral analysis!  Spatial analysis might
be necessary for localizing the burst.  All the counts
within a PSF radius and within a time range can be binned
into a count spectrum (apparent energy is the single
dimension), and traditional spectral analysis can be
applied to the resulting series of LAT count spectra.  The
GBM data (also a list of counts) can be binned with the
same time binning, and then joint fits can be performed.

The afterglow will most likely produce a small number of
counts accumulated over timescales of tens of minutes to
hours. Thus afterglow data must be analyzed with the
general likelihood tool being developed for LAT data
analysis.

The LAT team and the GLAST Science Support Center (GSSC)
are developing a suite of tools to analyze both LAT and GBM
data.  These tools will use the HEAdas system supported on
both Windows and LINUX platforms; most of the tools will be
FTOOLS.  Therefore the data will be in FITS files, and the
tools will use IRAF-style parameter files. Here we describe
the methodology for burst spectral analysis.

{\bf Extract LAT Counts:} The user will extract the LAT
counts from a specified time and region (here a circle
around the burst position) from a GSSC database.  The
web-based extraction tool will return a FITS file with the
requested counts and a second FITS file describing the
instrument's pointing and livetime during the burst.  Users
will have a tool to perform further selections.

{\bf Extract GBM Counts:} The GSSC will provide GBM counts
in a FITS file.  Users will fit polynomial (in time)
background models to data before and after the burst.

{\bf Bin Counts:} An event binning tool will bin the LAT or
GBM counts in both energy and time, resulting in a series
of count spectra spanning the burst.  The energy grid will
be user specified.  The time bins may be a)~equally spaced,
b)~user specified, c)~chosen to have equal signal-to-noise
ratios in a user-specified energy band, or d)~chosen by
Bayesian Blocks\cite{3}.  The LAT counts are assumed to
have no background, while the expected GBM background in
each bin will be calculated from the background model.

{\bf DRM Generation:} The LAT response function will be
collapsed into a Detector Response Matrix (DRM), the
product of the effective area at the position of the burst
and the energy redistribution matrix.  The effective area
will account for the size of the region from which the
counts were extracted. GBM DRMs will be supplied for each
burst, and users will have a tool to calculate their own
GBM DRMs.

{\bf Spectral Fitting---Binned Spectra:} The spectra can be
fit using XSPEC, with scripts automating the fitting of
series of spectra. Spectra from the LAT, GBM and other
missions (e.g., Swift) may be fit separately or jointly. Of
course, the relative calibration of the different
instruments will have to be understood.  The XSPEC team is
adding the capability of saving the results of the spectral
fitting, along with the model spectra.

{\bf Spectral Fitting---Spectra Unbinned in Energy:} For a
burst with few counts a likelihood treatment using a
variant of the standard likelihood tool may be more
powerful; see Fig.~1.  In this case the likelihood function
will be calculated for the apparent energy of LAT counts
accumulated over a time period of interest. Assuming the
LAT spectrum is an extrapolation to higher energy of the
GBM-observed spectrum, the GBM spectral fits can be used as
priors on the LAT parameters.
\begin{figure}
  \includegraphics[height=.3\textheight]{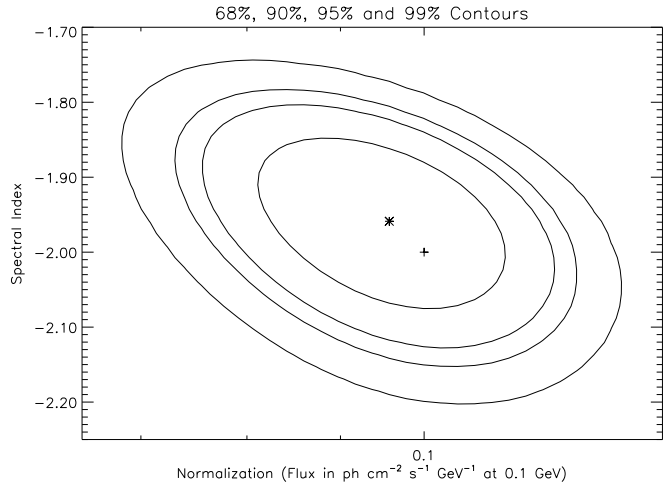}
  \caption{Confidence region for a sample power law LAT count
spectrum with 115 counts.  The asterisk marks the
likelihood maximum and the cross the input parameters. }
\end{figure}

\end{document}